\documentclass[twocolumn,aps,pre,amsmath,showpacs]{revtex4}
\usepackage{graphicx}
\usepackage{dcolumn}
\usepackage{bm}
\usepackage{amssymb}
\usepackage{amsmath}
\usepackage{color}

\def\be{\begin{equation}}
\def\ee{\end{equation}}

\begin{document}

\title{Theory of molecular crowding in Brownian hard-sphere liquids}
\author{Alessio Zaccone and Eugene M. Terentjev}
\affiliation{Cavendish Laboratory, University of Cambridge, JJ Thomson Avenue,
Cambridge CB3 0HE, U.K.}
\date{\today}
\begin{abstract}
We derive an analytical pair potential of mean force for Brownian molecules in
the liquid-state.
Our approach accounts for many-particle correlations of crowding particles of
the liquid, and for diffusive transport across the spatially modulated local
density of crowders in the dense environment. Specializing on the limit of
equal-size particles, we show that this diffusive transport leads to additional
density- and structure-dependent terms in the interaction potential, and to a
much stronger attraction (by a factor $\approx 4$ at average volume fraction of
crowders $\phi_0=0.25$) than in the standard depletion interaction where the
diffusive effects are neglected. As an illustration of the theory, we use it to
study the size of a polymer chain in a solution of inert crowders. Even in the
case of athermal background solvent, when a classical chain should be fully
swollen, we find a sharp coil-globule transition of the ideal chain collapsing
at a critical value of the crowder volume fraction $\phi_c \approx 0.145$.
\end{abstract}
\maketitle

\section{Introduction}
The theory of the liquid state relies heavily on the statistical mechanics of
hard-sphere systems~\cite{Hansen}. One of the most remarkable among these
results is the Percus-Yevick integral equation, the solution of which yields an
accurate description of the structure of hard-sphere liquids in terms of the
radial distribution function $g(r)$. The latter is related to the probability
of finding a second particle at a position $r$ along the outward radial
coordinate measured from the test particle. $g(r)$ typically features a peak
near contact (the first coordination shell) associated with an enhanced
probability of finding a second particle near contact. This is due to the
osmotic pressure exerted by all the other particles in the system which is not
balanced in the gap between the two particles near contact, an effect which is
known as \emph{depletion attraction}. This effect is clearly the result of
collective dynamics (``more is different'') since it introduces an attractive
interaction which depends on the density, and is not present originally in the
pair interaction between two hard particles. In ensembles of hard spheres of
equal size, this attraction induced by many-body effects is known to be weak:
even at high density it is barely of the order of the thermal energy scale
$kT$~\cite{dijkstra}. This attraction, however, may become substantial in
binary mixtures of hard spheres where spheres of two different sizes coexist:
this is an important attractive interaction that determines the phase behavior
of colloid-polymer mixtures~\cite{frenkel} and its first theoretical
description (valid for dilute systems) was given long ago in a famous work by
Asakura and Oosawa~\cite{oosawa}. This attraction plays an important role in
biological systems where, e.g., it is a factor controlling cellular
organization~\cite{marenduzzo}.

However, in both colloidal and biological systems, the particles are not simply
hard spheres subject to Newtonian dynamics. Instead, they are Brownian
particles obeying Langevin dynamics and they are sensitive to the local
chemical potential of solvent~\cite{Dhont}. Since this aspect has been
neglected both in theoretical and simulation studies of depletion in the past,
it is an interesting question to ask: what is the effect of the dense solvent
and the diffusive nature of particle dynamics on the depletion attraction. Here
we show that many-particle correlations that lead to the modulated radial
distribution $g(r)$ can strongly affect the local osmotic pressure profile
between two test particles. As a result, diffusive driving forces arise which
push the two particles together more strongly than if the dynamics were purely
Newtonian, as has been assumed in \cite{oosawa} and ever since. This is
reflected in a much stronger depletion attraction for Brownian particles as
compared to hard spheres in vacuo, such that the attraction minimum can be
substantial also in systems of equal-size particles.

We then apply this theory to calculate the coil-globule transition of ideal
polymer chains in solution with Brownian particles of
approximately the same size as the monomers.
It was observed 40 years ago that
adding polyethylene glycol (PEG)
molecules to solutions of DNA in water leads to the collapse or condensation of
DNA as soon as a critical concentration of PEG
is reached~\cite{Lerman}. Several theories have been proposed to explain this
``crowding effect'',
which has widespread implications as the biological function of all biopolymers
is strongly dependent on their size and conformation, in
particular, on whether they are in a swollen or collapsed
state~\cite{Bloomfield1}.  This stark distinction in biological
function is reflected in the names of their ``denatured'' and ``native''
states of biopolymer molecules. One of the most familiar cases is the packing
of DNA,
in dense cellular environment, and in vitro experiments where the crowding
was artificially tested~\cite{DNA,Lavrentovich}. In proteins subtle changes of
molecular
environment, instead of natural folding, could trigger their aggregation into
toxic assemblies (such as the amyloid fibrils responsible for
neuro-degenerative diseases)~\cite{Knowles}. It has been shown that folding
can be promoted by the addition of inert crowders via a mechanism which
bears many similarities with the crowding-induced globular collapse of generic
polymers~\cite{Thirumalai}.

The effective interaction between pairs
of polymer segments in the presence of quenched randomly distributed impurities is a classical problem in polymer physics, originally solved by Edwards, Muthukumar and Baumgaertner~\cite{Edwards, Baum}. When the impurities are mobile (annealed), a similar attraction
also exists and is likewise due to entropic effects, this time in the form of depletion forces~\cite{Thiru}. Despite the similarities, the mobile crowding problem has
a different formulation because the effective potential in this case has to be related
to the whole diffusive processes acting between the polymer segments and the Brownian crowders. The interplay between diffusion in a dense
environment with a liquid structure and the depletion effect, is the focus of our work.

The theories proposed to explain this phenomenon attempt to account for the
many-particle correlations due to crowders,
which induce an effective attraction between pairs of monomers in the
chain~\cite{Vrij,grosberg,Sear,Schoot,gelbart,szleifer}. This is reflected in a
decreasing second virial coefficient upon increasing the volume fraction of
crowders in the system, until the second virial coefficient finally changes
sign, from positive to negative, and a transition from coil to globule of the
polymer occurs~\cite{gelbart}. None of these theories, however, account for the
role of the microscopic liquid structure in this process and they all
generically attribute the increased effective attraction to the increased
\emph{global} (uniform) osmotic pressure exerted by the crowding particles.
These approaches also neglect the role of the diffusive dynamics. In contrast,
our approach uses as its input the radial distribution function of the fluid
$g(r)$, which develops a modulated structure at increased densities. It thus
allows us to obtain additional forces generated by diffusive transport across
\emph{local} gradients of the osmotic pressure (and the corresponding chemical
potential). At higher densities this effect turns out to be much stronger than
the standard depletion forces, and it of course vanishes in the low-density
limit where one recovers the standard depletion attraction given, for equal
size spheres, by $-\ln g(r)$. The fact that all depletion-based theories
underestimate the crowding effect manifests itself, for instance, in that the
critical density of crowders for the polymer collapse, derived
in~\cite{Schoot}, diverges to infinity when the size of the crowder particle
matches the size of the polymer Kuhn length -- clearly an unphysical outcome.

There are several steps in our development:
(1) We first derive the analytical form of this effective pair
potential between monomers, by accounting for multi-particle correlations and
also for the presence of the solvent, which shows an attractive well,
increasing at higher concentration of crowders. (2) This effective potential
requires an expression for the radial distribution function $g(r)$, which we
calculate using the currently most accurate analytical approach based on the
Wertheim solution of the Percus-Yevick equations. (3) Using this effective
potential we then calculate the second virial coefficient for monomer-monomer
interaction of a polymer chain in good solvent, and use a simple
scheme of Flory mean-field free energy to qualitatively describe the chain
collapse at a
critical volume fraction of crowders.
Our theory uses only one serious assumption, the local density
approximation (LDA), the validity of which is carefully examined below.
Therefore,
this analysis can be applied to any kind of system, provided that the structure
factor of the monomer-crowder mixture can be extracted from e.g. radiation
scattering experiments.

\section{{Derivation}}

Consider a Brownian particle moving in a sea of other similar Brownian
particles, which
all mutually interact via excluded volume (purely hard-sphere particles). We
let the particles be embedded in
a background solvent, which is a simple liquid providing the thermal bath. Let
us take some arbitrarily chosen point
of the system as the origin and be $\bm{r}$ the distance vector of the
tagged particle from the origin. In the local density approximation
(LDA)~\cite{Hansen}, let $\mu_{s}(\bm{r})$ and $\mu_{B}(\bm{r})$ be the
chemical potentials of the solvent and of the Brownian particle, respectively.
Then the local driving forces acting on a solvent molecule and on a Brownian
particle are given by $\bm{F}_{s}=-\nabla_{}\mu_{s}(\bm{r})$ and
$\bm{F}_{B}=-\nabla_{}\mu_{B}(\bm{r})$, respectively. {By its definition, the
spatial variation of chemical potential is related to the variation in the free
density per particle associated with the fact that both the Brownian particles
and the background solvent molecules have a locally higher free energy in
regions where the concentration of Brownian molecules or, respectively, solvent
molecules, is higher. As a consequence a driving force caused by the gradient
of chemical potential acts to bring Brownian molecules from region where their
concentration is higher into regions where their concentration is lower, and a
similar driving force acts on the solvent molecules. At steady state, the two
forces must be separately equal to zero.}
In fact, the two chemical
potentials at constant pressure and temperature are related by the Gibbs-Duhem
equation: $\rho(\bm{r})\nabla_{}\mu_{B}(\bm{r}) +
\rho_{s}(\bm{r})\nabla_{}\mu_{s}(\bm{r})=0$, where $\rho(\bm{r})$
is the local density of Brownian particles and $\rho_{s}(\bm{r})$ is the
local density of the solvent. Further, the mass conservation relation is valid:
$v_{B}\rho(\bm{r})+v_{s}\rho_{s}(\bm{r})=1$, where $v_{B}$ and
$v_{s}$ are the volumes of the Brownian particle and of the solvent molecule,
respectively. The relevant, net force which acts on the Brownian particle is
{the one \emph{relative} to the solvent}, which is obtained by subtracting the
force locally experienced by the background solvent. {In order to ensure that
the unit volume of background solvent is at rest with respect to the particle,
we need to consider forces per unit volume instead of the bare
forces~\cite{Dhont}.}
Hence we have: $\bm{F}/v_{B}=\bm{F}_{B}/v_{B}-\bm{F}_{s}/v_{s}$, and then the
net force acting on a single particle is given by:
$\bm{F}=\bm{F}_{B}-(v_{B}/v_{s})\bm{F}_{s}$. Substituting $v_s$ and $\rho_s$,
after a simple algebra we obtain the following general expression
for the net force acting on a Brownian particle moving in the solvent populated
with other Brownian particles:
\begin{equation}
\bm{F}=-\frac{1}{\rho(\bm{r})}\nabla \Pi(\rho(\bm{r})) ,
\label{eq:force}
\end{equation}
where the local osmotic pressure is defined as: $\Pi(\rho(\bm{r}))\equiv
-(\mu_{s}(\bm{r})-\mu_{s}^{0})/v_{s}$, with $\mu_{s}^{0}$ the constant chemical
potential of the pure solvent that does not contain any Brownian particles.
This expression for the force is a consequence of our using the LDA and is only
valid when the local variation in density is
smooth, that is $\nabla \rho /\rho \ll 1/\xi$ with $\xi$ the characteristic
range of oscillations of the radial distribution function $g(r)$
\cite{Barrat,Hansen}. It is clear, e.g. from the Fig.~\ref{fig2}, that the
interesting
effects of crowding-induced attraction may occur at sufficiently low densities
where this LDA condition is satisfied.

Equation (\ref{eq:force}) can also be derived from dimensional arguments. In a
suspension of Brownian
particles, $\Pi$ is the pressure exerted by the suspension itself. Then
$-\nabla \Pi$ represents the local density of force acting
in the suspension~\cite{zeldovich}. If $\rho$ is the number density of
particles, it follows that the force acting on average upon a single particle
of the suspension is given by: $-(1/\rho)\nabla \Pi$. This force is zero
whenever the system is spatially homogeneous, as it ought to be, e.g. for
infinitely diluted suspensions in the ideal-gas regime. If, however, the
distribution of the particles is not homogeneous, this diffusive force acts to
spread out local concentration gradients. As a result of this force, the
particle acquires a velocity which points along the direction of local steepest
decrease of the osmotic pressure.
 The close connection with diffusive motion becomes evident upon associating it
with the diffusive flux  $\bm{j}=\rho \bm{u}$, where the overdamped particle
velocity is: $\bm{u}=({D}/{kT})\bm{F}$. By applying the continuity operator to
this flux: $\dot{\rho}=-\nabla \cdot \bm{j}$, in the low-density limit of
$\Pi=kT\rho$, one recovers the Fickian diffusion equation:
$\dot{\rho}=D\nabla^{2}\rho$. Within LDA, one replaces
 \begin{equation}
\nabla\Pi=\nabla \rho \cdot \frac{\partial\Pi}{\partial \rho} ,
 \end{equation}
thus obtaining the diffusion equation valid for non-dilute and locally
inhomogeneous systems:
 \begin{equation}
 \frac{\partial\rho(\bm{r})}{\partial t}=D_{c}(\rho)\nabla^{2}\rho(\bm{r}).
  \end{equation}
The \emph{collective} diffusion coefficient $D_{c}=({D}/{kT})
{\partial\Pi}/{\partial \rho}$ takes care of the increased density
effects~\cite{Dhont}.
This derivation highlights the fact that Eq.(\ref{eq:force}) represents the
driving force acting on a Brownian particle in non-dilute and locally
inhomogeneous systems, and is consistent with the particle migration being
governed by diffusion, which in turn is driven by gradients of the chemical
potential.

\begin{figure} 
\includegraphics[width=.85\linewidth]{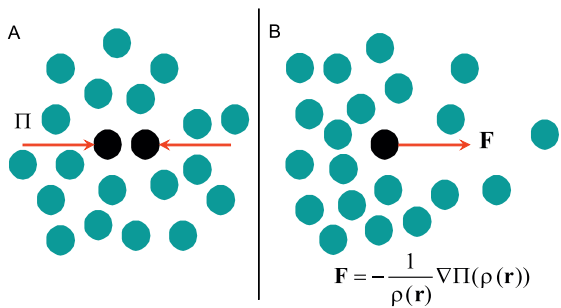}
\caption{(Color online) A: The classical depletion effect when no third particle can be
accommodated in the gap between the two tagged particles.
As a consequence, the particles are pushed towards each other by a force
proportional to the {global} osmotic pressure of the suspension $\Pi$ which
remains unbalanced in the gap. B: The generic effect of the diffusive driving
force, Eq.(\ref{eq:force}), proportional to the {local} gradient of the osmotic
pressure and driving the particle down the concentration gradient. This force
must also be active on the two test particles of panel A, because upon moving
closer together they are effectively moving across a local concentration
gradient of a dense fluid of crowders.}
\label{fig1}
\end{figure}

\begin{figure} 
\includegraphics[width=.85\linewidth]{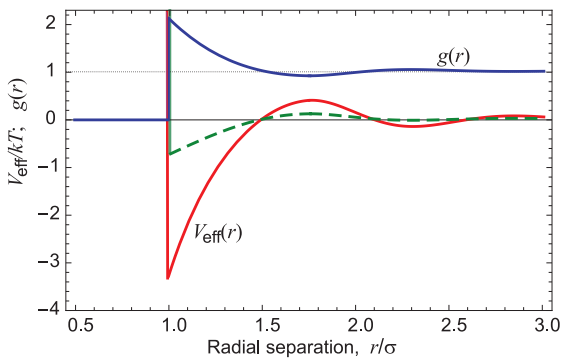}
\caption{(Color online) Radial distribution function and effective pair potential for
$\phi_0=0.25$ calculated using the Percus-Yevick theory~\cite{Henderson}. There
are no free parameters in this athermal system with only excluded volume
between particles in contact. The dashed line shows the prediction of the
depletion
theory, $-\ln g(r)$, for comparison.}
\label{fig2}
\end{figure}

Let us now consider that the center of frame from which $\textbf{r}$ is
measured coincides with a test particle and consider a second particle
approaching the test particle, in the
moving spherical coordinates frame centered on the test particle. The work
required to bring the second
particle to a point $\bm{r}$ measured in the moving frame of the first particle
is given by
\begin{equation}
W=-\int\bm{F}\cdot d\bm{r}
=\int_{\infty}^{r}\frac{1}{\rho(r)}\frac{d}{dr}\Pi(\rho(r))dr .
\label{eq:work}
\end{equation}
In the second equality we used Eq.(\ref{eq:force}) and the fact that with
hard-spheres the distribution of particles is spatially isotropic, thus only
the radial distance $r=|\bm{r}|$ between the two monomers is a variable of
the effective pair-potential energy $V_{\mathrm{eff}}(r) \equiv W$ acting
between the two Brownian monomers. To evaluate $V_{\mathrm{eff}}(r)$ we need an
equation of
state for the crowders to relate the osmotic pressure and the density, and we
will assume that this relation also applies \emph{locally} which, once more,
amounts to making use of LDA. For the equation of state
we can use the Carnahan-Starling (CS) equation of state which is accurate up to
quite high liquid densities~\cite{Hansen} and we write it as:
\begin{equation}
\frac{1}{kT}\frac{\Pi(\rho(r))}{\rho(r)}=\frac{1+\phi(r)+\phi(r)^2-\phi(r)^3}{(1-\phi(r))^3}
\label{eq:carnahan}
\end{equation}
where $\phi(r)=(\pi \sigma^{3}/6)\cdot \rho(r)$ is the volume (packing)
fraction of
crowders ($\sigma$ chosen as the diameter of the Brownian particle). The CS
equation of state is frequently
used in standard approaches for the calculation of depletion forces. However,
we should remark that
in those cases the aim is to account for density effects on the free energy of
interaction
which for Newtonian hard-spheres is a purely mechanical effect. That is, one
takes the work $W\propto\Pi\cdot V_{d}$, with $V_{d}$ is the depletion volume
(a function of $\bm{r}$), and the CS equation is used to relate the global
($\bm{r}$-independent) osmotic pressure of the solution $\Pi$ to the overall
density of the crowders $\phi_{0}$. In those approaches $\Pi$ and $\phi$ are
not allowed to vary with $\textbf{r}$ and therefore the driving force in
Eq.(\ref{eq:force}) due to {local} variation in particle density is neglected.
See Fig.~\ref{fig1} for a visual explanation of this contrast.

Using this CS expression (\ref{eq:carnahan}) in the r.h.s. of
Eq.~(\ref{eq:work}) and integrating we obtain
\begin{equation}
\frac{V_{\mathrm{eff}}(r)}{kT}=-\frac{\phi(r)-3}{(\phi(r)-1)^3}+\frac{\phi(r)-3}{(\phi_0-1)^3}
-\ln\frac{\phi(r)}{\phi_0},
\label{eq:veff}
\end{equation}
where it is important to distinguish the local packing fraction of crowders
$\phi(r)$ from the mean fraction $\phi_0$ at $r
\rightarrow \infty$. By construction, this is an effective pair potential of
interaction between two
Brownian particles. In addition to the unbalanced osmotic pressure in the
depletion gap near contact (which is expressed by the standard $-\ln [\phi(r) /
\phi_0] \equiv -\ln g(r)$ term), it accounts also for the diffusive drift
pushing the two particles against each other because of the local liquid
structure, expressed by the two fractions in Eq.(\ref{eq:veff}).

Equation~(\ref{eq:veff}) is the key result of this work. Note that
the $-\ln g(r)$ term represents the depletion effect, which many of the earlier
important publications have investigated
thoroughly~\cite{oosawa,Sear,Thirumalai}.
For dilute systems we have that the radial density profile around the monomer
tends to become flat, $\phi(r)\rightarrow \phi_0$ for any $r$. In this limit,
one
can readily check via de L'H$\hat{\mathrm{o}}$pital's rule that:
$\lim_{\phi \rightarrow 0}V_{\mathrm{eff}}=- kT \ln g(r).$
This limit is a well-known law of statistical mechanics~\cite{Hansen} which
holds in
the dilute ideal-gas limit independently of the particular form of the
pair-potential.
From another perspective, since this theory describes the effective pair
potential
between particles of a ``Brownian fluid'', this result means
that in the dilute limit it does not matter whether the particles are Brownian
or not, which is an important observation on its own. Further, our result
extends the fundamental potential of mean force from the standard simple fluids
to the Brownian fluids.

\begin{figure} 
\includegraphics[width=.85\linewidth]{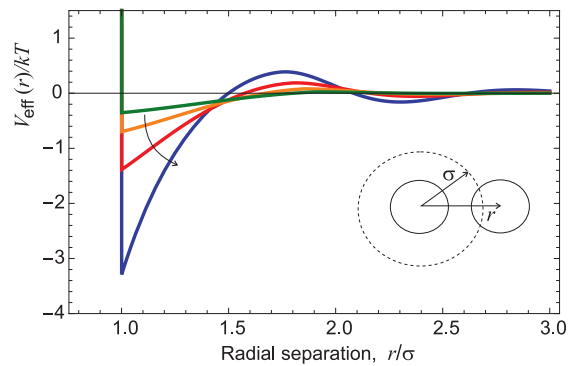}
\caption{(Color online) The effective potential upon increasing the mean volume fraction
of Brownian crowders, $\phi_0 = 0.1, \ 0.15, \ 0.2$ and $0.25$ for the
successive
curves.}
\label{fig3}
\end{figure}

As we have explained above, depletion alone underestimates the effective
attraction in Brownian systems, which is explicitly illustrated by plotting the
two forms of effective potential in Fig.~\ref{fig2}, after substituting the
explicit form of $g(r)$ at a given mean density $\phi_0$ as explained in the
Appendix~\cite{Henderson}. This effective pair potential is plotted for
$\phi_0=0.25$, which is a relatively high
density still well within the fluid region of the phase diagram of hard
spheres.  It is evident
that the effective potential develops a short-range attractive well aligned
with
the first coordination shell of the radial distribution function. It is also
clear that the
attraction minimum is significantly deeper, by a factor $\simeq 4$, than the
one given by the
standard depletion potential $-\ln g(r)$ alone which suffices to describe hard
spheres in vacuo
(that is, ignoring the diffusion of particles in the suspending matrix). The
two new terms in the
 r.h.s. of Eq.(\ref{eq:veff}) turn out to be a major effect not captured by
earlier theories.

Figure~\ref{fig3} plots the evolution of the effective potential
$V_{\mathrm{eff}}(r)$ on increasing the mean concentration of crowders in
solution, illustrating that the depth of the attractive well exceeds $kT$ at
$\phi_0 \geq 0.15$. It is interesting to note that also the local maximum, the
energy barrier in correspondence of the local minimum  in the $g(r)$, is
significantly larger for Brownian particles than for Newtonian hard spheres.
This can be understood by recalling that the diffusive driving force acts to
push the particles down the $\rho$ and $\Pi$ gradients. Upon approaching each
other the two particles are separated by a (spherical) ``monolayer'' of third
particles of the first coordination shell. In order to come closer together,
the two particles need to
diffuse ``uphill'', i.e. jump over the particle in between. In this case the
diffusive driving force is pointing outwards along $r$ and it generates an
additional repulsion with respect to the bare local osmotic pressure which is
reflected in the higher local maximum of $V_{\mathrm{eff}}$
as compared to the case of hard spheres in vacuo. This effect is
particularly important for diffusion-limited reaction kinetics in crowded
systems~\cite{foffi,McCammon,Zaccone}).

\section{{Coil-globule transition driven by crowding}}

\begin{figure} 
\centering
\includegraphics[width=.75\linewidth]{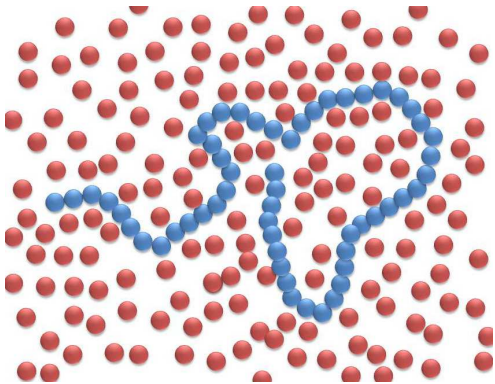}
\caption{(Color online) Schematic representation of the model used: an
excluded-volume chain in a suspension of Brownian particles. The background
solvent, taken to be structureless, is not shown.}
\label{fig4}
\end{figure}

\begin{figure} 
\includegraphics[width=.75\linewidth]{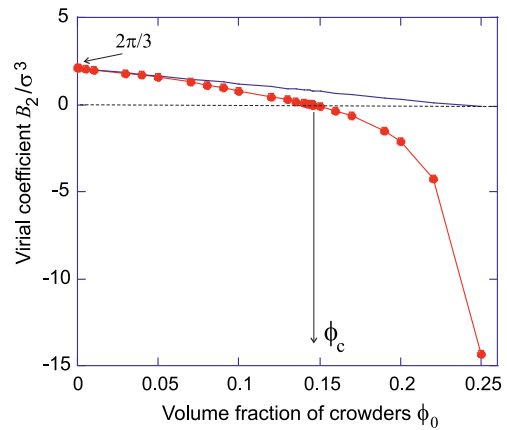}
\caption{(Color online) The values of $B_2$ calculated numerically from
Eq.(\ref{eq:b2}), showing the point of changing sign at a
critical concentration $\phi_c \approx 0.145$. The
straight line illustrates the linear-order approximation,
$B_2 \approx (2\pi \sigma^3 /3)(1-(17/4)\phi_0)$, which strongly
overestimates the critical point $\phi_c$.}
\label{fig5}
\end{figure}

Let us now imagine that the two interacting particles represent two monomers of
a polymer chain in a
suspension of other Brownian molecules (crowders) which interact with the two
test monomers via excluded volume. The model is sketched in Fig.~\ref{fig4}.
For this system, we note that the effective interaction derived above is a
function of the overall density in the system, which
includes both the monomers and the crowders. In a more precise way we could
decompose the two contributions such that:
\begin{equation}
g(r) \equiv  \frac{\phi(r)}{\phi_0} = \frac{\phi_{mc}(r)}{\phi_{\rm tot}} +
\frac{\phi_{mm}(r)}{\phi_{\rm tot}},
\end{equation}
where $\phi_{mc}$ is the local profile of crowders around the test monomer,
$\phi_{mm}$ is that of other monomers, and $\phi_{\rm tot}$ is the overall bulk
packing fraction. At high density of crowders, $\phi_{mc} \gg \phi_{mm}$ and
the general radial distribution function is determined by crowders only $g(r)
\approx \phi_{mc}(r)/\phi_0$, while in the opposite limit of $\phi_{mc}
\rightarrow 0$ we have $g(r) \approx \phi_{mm}(r)/\phi_0$, that is, only
determined by the hard-sphere interaction of monomers.
We can write down the effective second virial coefficient
$B_{2}$ for the two chosen polymer segments. Normally one expects $B_2$ to be a
function
of $T$ (as well as the mean concentration $\phi_0$), however, in our case
this expression is purely entropic and therefore completely athermal:
\begin{equation}
B_{2}=2\pi\int_{0}^{\infty}r^{2}\left[1-e^{-V_{\mathrm{eff}}(r)/kT}\right]dr .
\label{eq:b2}
\end{equation}
The sign of the second virial coefficient controls whether the polymer chain
would be found in the collapsed state, which corresponds to $B_{2}<0$, or
swollen
($B_{2}>0$)~\cite{Rubinstein}, as we are going to discuss below.
For a given $\phi_0$ the sign of $B_{2}$ is dictated by the balance in the
integrand between the repulsive hard-core part of the effective potential (for
spheres of diameter $\sigma$: $B_2= 2\pi \sigma^3/3$) and
the attractive part due to the unbalanced osmotic pressure.
As the hard-core part is independent of $\phi_0$, the behavior of $B_{2}$ as a
function of the volume fraction of crowders is therefore controlled by
the behavior of the attractive part of $V_{\mathrm{eff}}(r)$ as a function of
$\phi_0$. As shown in Fig.~\ref{fig3}, the strength of the attractive part of
$V_{\mathrm{eff}}(r)$ is a monotonically increasing function of $\phi_0$. This
is intuitively expected because the unbalanced osmotic pressure in the excluded
volume gap between two monomers (the standard depletion) and the diffusive
driving force both become stronger with increasing density of crowders.
Figure~\ref{fig5} shows the
evolution of $B_2$ with increasing concentration of crowders, together with the
linear-order approximation valid at very low densities, which is easily
obtained by the series expansion of Eqs.(\ref{eq:veff}) and (\ref{eq:b2}).
There is a
remarkable similarity with the traditional $T$-dependence of $B_2(T)$ for the
Lennard-Jones type of attractive-repulsive potentials, however, we need to
emphasize once
again that our basic theory is purely entropic and therefore athermal. The
application of the effective
potential $V_{\mathrm{eff}}$ from Eq.(\ref{eq:veff}) to evaluate $B_2$, leads
to a change of sign
of the second virial coefficient in a
system of purely hard-core repulsive particles, when a critical
concentration $\phi_c \approx 0.145$ is reached.

As a simple exercise to illustrate the theory we take the Flory
excluded-volume theory of polymer chains, where the free energy is given by the
two contributions both determined by the size (radius of gyration $R$) of the
chain~\cite{Rubinstein}, the total potential energy of effective pair
interactions and the entropy of configurations of an ideal (Gaussian) random
walk:
\begin{equation}
F\simeq kT \left(B_{2}\frac{N^{2}}{R^{3}}+\frac{R^{2}}{Nb^{2}}\right),
\label{eq:flory}
\end{equation}
where  $N$ is the number of monomers in the chain and $b$ is the linear size of
the monomer. To simplify the key qualitative
results, let us assume $b=\sigma$, that is, the size of the chain segments and
the
crowders is the same, as illustrated in Fig.~\ref{fig4}. In spite of its many
conceptual shortcomings, the Flory expression (\ref{eq:flory})
is remarkably accurate. Although it neglects correlations between monomers
along the
chain, the errors in the internal energy and the entropic
terms are roughly the same, which leads to the cancelation of errors
responsible for the good accuracy of this crude model~\cite{Rubinstein}.
Minimizing
Eq.(\ref{eq:flory}) with respect to $R$ leads to the classical estimate of the
size of polymer chain swollen in good solvent:
\begin{equation}
R\simeq (B_{2}b^{2}N^{3})^{1/5}
\label{eq:r}
\end{equation}
which applies for $B_{2}\geq 0$. When $B_{2}<0$ the free energy
(\ref{eq:flory}) does not have a
well-defined minimum and the chain is considered collapsed into a dense
globule.
The theory of polymer collapse in a poor solvent is
developed \cite{vilgis,freed,three},
and it is known that to properly describe the globular (collapsed) state one
needs to invoke the third virial coefficient contribution to the interaction
energy, which provides the stabilizing effect against an apparent collapse to a
point,  $R \rightarrow 0$. In this paper we do not deal with the globular state
and the considerable computational difficulties associated with it. Our aim is
merely to illustrate the effect of molecular crowding on a chain in an
otherwise good
solvent, and describe the onset of the chain collapse, which occurs at
$B_{2}\geq 0$ where $B_3$ is not yet relevant.

\begin{figure} 
\includegraphics[width=.75\linewidth]{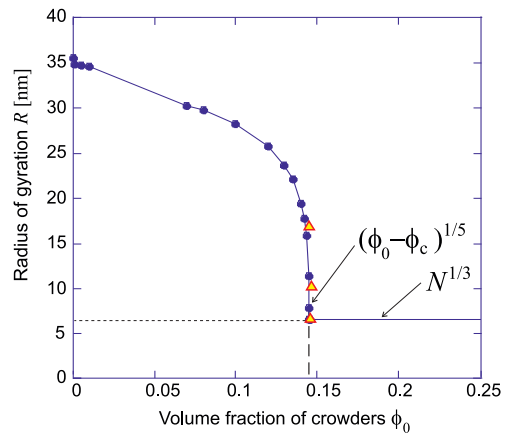}
\caption{(Color online) The size of the polymer as a function of the volume fraction of
molecular crowders for $b=\sigma=1nm$ and number of monomers $N=300$.
The critical transition occurs at the point $\phi_0 = \phi_c$ where the second
virial coefficient $B_2$ changes its sign. At high concentration the collapsed
polymer chain is assigned a dense-packed size $R\simeq bN^{1/3}$. The
additional data
points ($\triangle$) have been calculated using the perturbation expansion for
excluded volume. }
\label{fig6}
\end{figure}

Using Eq.(\ref{eq:veff}) for the effective potential evaluated making use of
Percus-Yevick
theory for $g(r)$~\cite{Henderson} in Eq.(\ref{eq:b2}) and the latter in
Eq.(\ref{eq:flory}), the polymer size as a function of the crowder
concentration
$\phi_0$ has been evaluated for $b=\sigma$ and plotted in Fig.~\ref{fig6}. The
power-law
scaling with the crowder density follows directly from Eq.(\ref{eq:r}) when one
substitutes
the linear relation $B_2 \approx {\rm const}\cdot (\phi_c-\phi_0)$ near the
transition point.

One may argue that the characteristic scaling of the critical point might
change with the third virial coefficient taken in to account. {Therefore, we
also made an estimate of the polymer size based
on the perturbative expansion by Muthukumar and Nickel~\cite{muthu}, which
accounts for the effect of the three-body collisions and for chain connectivity
in a simple way. The perturbation is done on the distribution function of the
end-to-end vector using $B_{2}$ as the small parameter near the point of its
changing sign. The full expression is reported as Eq.(2.100) in~\cite{doibook}
where, strictly, the perturbation expansion parameter is:
$z=(3/2\pi)^{3/2}B_{2}N^{1/2}/b^{3}$. A few points calculated at
$(\phi_0-\phi_c) \rightarrow 0$ with this approach are shown in Fig.~\ref{fig6}
and are found to effectively coincide with the more naive Flory calculation.
Hence we will stick to the handier Flory description to analyze the physics
predicted by the model.}

The theory predicts a sharp collapse of the polymer at a well-defined
critical value of the crowder volume fraction, at $\phi_{c}\simeq 0.145$,
which is the point at which $B_{2}$ changes sign and
becomes negative. This value is close to the result of~\cite{Khokhlov}
where it was experimentally found that DNA, under conditions of screened
electrostatics, collapses sharply
when 10-20\% of PEG is added to the solution. In the limit of infinite dilution
of crowders,
$\phi_0=0$, the effective potential for the monomers reduces to the pure
hard-core interaction and $B_{2}=2 \pi b^{3}/3$ (since we treat the
background solvent as a structureless continuum) and thus we recover the limit
of fully swollen self-avoiding chain in athermal solvents: $R = (2\pi /3) b
N^{3/5}$.
Previous approaches treat the crowders like Newtonian hard spheres with a
uniform local density
and, therefore, underestimate the attraction. In this context, the theory
of~\cite{gelbart} predicts no collapse for
systems where crowders and monomers are comparable in size, contrary to the
evidence of
collapse also in that limit from experiments~\cite{Khokhlov}.

In the present study we have focused, to preserve analytical clarity of the key
points, on crowding
molecules which have a size comparable to the size of the monomers in the
polymer chain.
In future studies, this theory should be extended to arbitrary monomer/crowder
size ratios.
The theory in fact accounts for this ratio through the radial distribution
function $g(r)$ of
crowders around a monomer, which can be extended to the case of arbitrary size
ratios by means of a theory by Lebowitz~\cite{Lebowitz}. Unfortunately, this
can be done only numerically.
However, we can anticipate that, upon increasing the size of the
crowders, the packing around the monomer becomes less efficient and the first
peak in the $g(r)$ gets smaller at the same mean packing fraction
$\phi_0$~\cite{Lebowitz,Hansen}. This
means that the collapse transition occurs at higher $\phi_0$ upon increasing
the
crowders size with respect to the monomer. The reverse occurs upon decreasing
the crowder/monomer size ratio: the first peak is higher at comparatively lower
$\phi_0$ and therefore the collapse occurs at lower $\phi_0$ too.
Another interesting direction would be to extend this theory to specific
practically
relevant biomolecular systems by simply adding the specific interactions
(hydrophobic, hydrogen-bonding, etc, as well as the chain bending rigidity)
that are relevant in each case, e.g. in packing the DNA or (mis)folding
proteins. These additional potentials act as perturbations by shifting the
critical concentration of the coil-globule collapse, but the effect of crowding
is still
controlled by the effective two-particle potential  $V_{\rm eff}(r)$, which is
essentially entropic, excluded-volume driven. Previous
approaches~\cite{gelbart} omit the microscopic
details of the local structure induced by the crowders and therefore are not
suitable for
a realistic description of the interaction chemistry beyond the hard-sphere
interaction.

\section{{Conclusions}}
In summary, we have developed a theory that analytically describes the
effective interaction between two particles in Brownian hard-sphere liquids. In
addition to the multi-particle correlations which induce the depletion
attraction in hard-spheres in vacuo (governed by Newtonian dynamics), we also
account for the diffusive driving force acting on the two Brownian particles
due to the local gradients of osmotic pressure and chemical potential dictated
by the liquid-like structure. In fact, when two particles move into the
depletion gap (near-contact region) they are subject not only to the unbalanced
osmotic pressure in the gap (the cause of the standard depletion attraction),
but also to the additional diffusive driving force which pushes them towards
low-density regions (the depletion gap) thus enhancing the depletion
attraction. Slightly further apart, the particles also experience a much higher
repulsive barrier where the local density decreases between the first and the
second coordination spheres. The effective potential $V_{\rm eff}(r)$ uses the
structure factor $g(r)$ to calculate the forces between the pair of particles
mediated by the structured dense fluid dominated by Brownian motion.

It turns out that this additional effect can be very substantial in systems of
equal-size particles where it increases the attraction by over a factor of $4$
(at $\phi_0=0.25$) in comparison to the standard depletion interaction term
$-\ln g(r)$ for hard spheres in vacuo. The latter form is correctly recovered
in the low density limit $\phi_0 \rightarrow 0$, where the local gradients of
density and osmotic pressure disappear and there is no difference between hard
spheres in vacuo and Brownian hard spheres.

As an illustration of the theory, we applied it to calculate the size of a
polymer chain in solution with Brownian particles (crowders) of size equal or
comparable to the monomers. Our theory predicts a sharp coil-globule transition
occurring at a packing fraction of crowders $\phi_c \approx 0.145$, which is
the volume fraction at which the second virial coefficient of the monomers
changes sign owing to the effective pair potential. This approach is fully
microscopic because it uses as input the radial distribution function of the
monomer/crowder system which is potentially capable of accounting for
non-trivial chemistry details, whereas earlier approaches to depletion cannot
go beyond the hard-sphere interaction between monomer and crowder. The enhanced
attraction due to the diffusive drift in a structured environment of crowders
leads to the collapse of the polymer, in agreement with simulation and
experimental results, whereas previous studies neglecting the diffusive effect
(and thus underestimating the attraction) predict that collapse cannot occur in
that limit.

\subsection*{Acknowledgments}
This work has been supported by the EPSRC TCM Programme grant and by the Ernest
Oppenheimer Fellowship.

\appendix
\section*{APPENDIX: ANALYTICAL FORM OF $g(r)$}

 In order to calculate the potential of mean force between monomers, we need to
evaluate the radial distribution function $g(r)$. It is possible to do so
analytically for the case of spherical crowders which have size comparable to
the monomers and interact purely by excluded volume. In this case the $g(r)$ is
the one of hard-sphere fluids at given density $\rho\sigma^{3}$ and volume
fraction $\phi_{0}=\pi\rho\sigma^{3}/6$. The following analytical representation of
$g(r)$ has been used for our calculations, which is based on the analytical
solution of the Percus-Yevick equations by Wertheim~\cite{Wertheim} and the
successive improved parameterizations by Trokhymchuk, Nezbeda, Jirsak and
Henderson~\cite{Henderson}. According to the work of ~\cite{Henderson}, the
$g(r)$ can be constructed by splitting it into a depletion part valid for
$\sigma<r<r^{*}$ and a structural part valid for
$r>r^{*}$. The depletion part describes the first coordination shell up to the
first (depletion) minimum whereas the structural part describes the damped
oscillations at farther distance. The analytical expressions read as
follows~\cite{Henderson,TrokhyErr}:
\begin{equation}
\begin{aligned}
g(r)&=\frac{A}{r} e^{\mu (r - \sigma)} +
 \frac{B}{r} \cos[\beta (r - \sigma) + \gamma] e^{\alpha (r - \sigma)} \\
 &~\mathrm{for}~~\sigma<r<r^{*}\\
 &=1 + \frac{C}{r} \cos[\omega r + \delta] e^{-\kappa r}\\
 &~\mathrm{for} ~~r>r^{*}
\end{aligned}
\label{eqmod1}
\end{equation}
The parameters which figure in the above expressions are all ultimately
functions of $\phi_{0}$ only and can be evaluated using the improved
parameterizations of~\cite{Henderson} based on thermodynamic consistency
arguments. The parameters read as:
\begin{equation}
A = \sigma g_{\sigma} - B\cos\gamma
\label{eqmod2}
\end{equation}
\begin{equation}
B = \frac{g_{m} - (\sigma g_{\sigma}/r^{*}) e^{\mu (r^{*} - \sigma)}}
{\cos[\beta (r^{*} - \sigma) + \gamma] e^{\alpha (r^{*} - \sigma)} -
    \cos[\gamma] e^{\mu (r^{*} - \sigma)}} r^{*}
\label{eqmod3}
\end{equation}
\begin{equation}
C= \frac{r^{*} (g_{m} - 1) e^{\kappa r^{*}}}{\cos[\omega r^{*} + \delta]}
\label{eqmod4}
\end{equation}
\begin{equation}
\delta = -\omega r^{*} - \arctan[(\kappa r^{*} + 1)/(\omega r^{*})]
\label{eqmod5}
\end{equation}
where
\begin{equation}
g_{\sigma} = \frac{1}{4 \phi_{0}} \left(\frac{
     1 + \phi_{0} + \phi_{0}^2 - 2/3 \phi_{0}^3 - 2/3 \phi_{0}^4}{(1 - \phi_{0})^3}-1\right)
\label{eqmod6}
\end{equation}
\begin{equation}
\begin{aligned}
g_{m} = &1.0286 - 0.6095 \phi_{0} + 3.5781 \phi_{0}^2 - 21.3651 \phi_{0}^3\\
   &+42.6344 \phi_{0}^4 - 33.8485 \phi_{0}^5
\end{aligned}
\label{eqmod7}
\end{equation}

\begin{equation}
r^{*}\sigma = 2.0116 - 1.0647 \phi_{0} + 0.0538 \phi_{0}^2
\label{eqmod8}
\end{equation}

\begin{equation}
\alpha\sigma = 44.554 + 79.868 \phi_{0} + 116.432 \phi_{0}^{2} - 44.652 e^{2 \phi_{0}}
\label{eqmod9}
\end{equation}

\begin{equation}
\beta\sigma = -5.022 + 5.857 \phi_{0} + 5.089 e^{-4 \phi_{0}}
\label{eqmod10}
\end{equation}

\begin{widetext}
\begin{equation}
\gamma = \arctan\left[-\frac{1}{
    \beta_{0}} \frac{\sigma(\alpha_{0}(\alpha_{0}^2 + \beta_{0}^2) - \mu
(\alpha_{0}^2 - \beta_{0}^2))(1 +
        \frac{1}{2}\phi_{0}) + (\alpha_{0}^2 + \beta_{0}^2 - \mu \alpha_{0})(1 + 2
\phi_{0})}{
   \sigma(\alpha_{0}^2 + \beta_{0}^2 - \mu\alpha_{0})(1 + \frac{1}{2} \phi_{0}) -
\mu(1 + 2 \phi_{0})}\right]
\label{eqmod11}
\end{equation}
\end{widetext}

\begin{equation}
\alpha_{0} \sigma= \frac{2 \phi_{0}}{1 - \phi_{0}}\left(-1 + \frac{d}{4 \phi_{0}} -
\frac{\phi_{0}}{2 d}\right)
\label{eqmod12}
\end{equation}

\begin{equation}
\beta_{0}\sigma = \frac{2 \phi_{0}}{1 - \phi_{0}}\sqrt{3} \left(-\frac{d}{4 \phi_{0}} -
\frac{\phi_{0}}{2 d}\right)
\label{eqmod13}
\end{equation}

\begin{equation}
\mu\sigma = \frac{2 \phi_{0}}{1 - \phi_{0}}\left(-1 - \frac{d}{2 \phi_{0}} +
\frac{\phi_{0}}{d}\right)
\label{eqmod14}
\end{equation}

\begin{equation}
d = [2 \phi_{0} (\phi_{0}^2 - 3 \phi_{0} - 3 + \sqrt{
      3 (\phi_{0}^4 - 2 \phi_{0}^3 + \phi_{0}^2 + 6 \phi_{0} + 3)})]^{1/3}
\label{eqmod15}
\end{equation}
Furthermore, the parameters $\kappa$ and $\omega$ have been evaluated according
to the parameterization of Roth et al.~\cite{Roth} and read as follows:
\begin{equation}
\kappa \sigma= 4.674 e^{-3.935 \phi_{0}} + 3.536 e^{-56.270 \phi_{0}}
\label{eqmod16}
\end{equation}
\begin{equation}
\omega\sigma = -0.682 e^{-24.696 \phi_{0}} + 4.720 + 4.450 \phi_{0}.
\label{eqmod17}
\end{equation}

\end{document}